\documentclass[aps,prl,reprint,amsmath,amssymb,floatfix,superscriptaddress]{revtex4-2}

\makeatletter
\let\frontmatter@footnote@produce\frontmatter@footnote@produce@endnote
\makeatother

\usepackage{float}
\usepackage{lipsum}
\usepackage{titlesec}
\usepackage{natbib}
\usepackage{graphicx}
\usepackage{physics}
\usepackage{bm}  
\usepackage{amsmath} 
\usepackage{braket} 
\usepackage[colorlinks,allcolors=blue,urlcolor=blue]{hyperref}

\expandafter\def\expandafter\normalsize\expandafter{%
    \normalsize%
    \setlength\abovedisplayskip{3pt}%
    \setlength\belowdisplayskip{5pt}%
    \setlength\abovedisplayshortskip{-8pt}%
    \setlength\belowdisplayshortskip{2pt}%
}

\begin{document}

\title{Magnetotransport in Topological Materials and Nonlinear Hall Effect\\ via First-Principles Electronic Interactions and Band Topology}

\author{Dhruv C. Desai}
\affiliation{Department of Applied Physics and Materials Science, and Department of Physics, \protect\\ California Institute of Technology, Pasadena, California 91125}
\author{Lauren A. Tan}
\affiliation{Department of Applied Physics and Materials Science, and Department of Physics, \protect\\ California Institute of Technology, Pasadena, California 91125}
\author{Jin-Jian Zhou}
\affiliation{School of Physics, Beijing Institute of Technology, Beijing 100081, China}
\author{Shiyu Peng }
\affiliation{Department of Applied Physics and Materials Science, and Department of Physics, \protect\\ California Institute of Technology, Pasadena, California 91125}
\author{Jinsoo Park}
\affiliation{Department of Applied Physics and Materials Science, and Department of Physics, \protect\\ California Institute of Technology, Pasadena, California 91125}
\affiliation{Department of Physics, Pohang University of Science and Technology, Pohang 37673, Korea}
\author{Marco Bernardi}
\email{bmarco@caltech.edu}
\affiliation{Department of Applied Physics and Materials Science, and Department of Physics, \protect\\ California Institute of Technology, Pasadena, California 91125}
%
\begin{abstract}
Topological effects arising from the Berry curvature lead to intriguing transport signatures in quantum materials. Two such phenomena are the chiral anomaly and nonlinear Hall effect (NLHE). 
A unified description of these transport regimes requires a quantitative treatment of both band topology and electron scattering. 
Here, we show accurate predictions of the magnetoresistance in topological semimetals and NLHE in noncentrosymmetric materials by solving the Boltzmann transport equation (BTE) with electron-phonon ($e$-ph) scattering and Berry curvature computed from first principles. 
We apply our method to study magnetotransport in a prototypical Weyl semimetal, TaAs, and the NLHE in strained monolayer WSe$_2$, bilayer WTe$_2$ and bulk BaMnSb$_2$. In TaAs, we find a chiral contribution to the magnetoconductance which is positive and increases with magnetic field, consistent with experiments.  
We show that $e$-ph interactions can significantly modify the Berry curvature dipole and its dependence on temperature and Fermi level, highlighting the \mbox{interplay} of band topology and electronic interactions in nonlinear transport. The computed nonlinear Hall response in BaMnSb$_2$ is in agreement with experiments. By adding the Berry curvature to first-principles transport calculations, our work advances the quantitative analysis of a wide range of linear and nonlinear transport phenomena in quantum materials.
\end{abstract} 
\maketitle
%
\titlespacing\subsection{0pt}{12pt plus 4pt minus 2pt}{0pt plus 2pt minus 2pt}

Band topology is a central theme in quantum materials and is responsible for emergent behaviors such as robust edge states~\cite{Kane_graphene_2005,Qi_TI_2011,Hasan_TI_2010}, topological superconductivity~\cite{Fu_superconducting_TI_2008}, spin Hall effects~\cite{Bernevig_hgte_2006,Qi_qshe_2010}, and unconventional transport regimes~\cite{Zhang_nonreciprocal_ti_2022,farrell_floquet_ti_2016,Hsieh_TI_2009}. This physics is relevant for wide-ranging applications in electronics, spintronics, and quantum technologies~\cite{He_TI_2019,He_TI_2022}. 
The theory of Berry curvature and topological invariants has provided a deeper understanding of topological phases~\cite{Hasan_TI_2010}. Recent advances have led to the prediction and discovery of topological transport regimes, including the chiral anomaly in topological semimetals~\cite{Armitage_review_2018,Zyuzin_chiral_2012,Ong_chiral_2021}, the nonlinear Hall effect (NLHE) in noncentrosymmetric materials~\cite{Du_NLHE_2021_review,Sodemann_NLHE_2015, suarez2025nonlinear}, and novel nonlinear transport regimes in magnetic systems~\cite{wang2023quantum, Xiao2021, Cano2024}. 
\\
\indent
Although predictions of chiral anomaly in solids have a long history~\cite{nielsen198}, this effect has been observed only recently in topological Weyl semimetals in the form of a large negative longitudinal magnetoresistance (LMR)~\cite{Xiong_Na3Bi_2015,Huang_taas_2015,Li_cd3as2_2015,Zhang_taas_2016,Liang_exptest_2018,Hirschberger_gdptbi_2016,Reis_chiral_2016,Li_chiral_2017}. The negative LMR is caused by charge pumping between two Weyl cones when parallel electric and magnetic fields are applied, and experiments on Weyl semimetals have achieved a decrease in resistivity by more than 50\% at large magnetic fields~\cite{Li_cd3as2_2015,Xiong_Na3Bi_2015}. This behavior contrasts with conventional magnetotransport in metals, where the LMR is positive and increases with magnetic field. 
The NLHE is a more recent development in topological transport and has become an exciting research topic, with measurements reported in a wide range of \mbox{materials}~\cite{Ma_NLHE_2019,Huang_2022_NLHE_wse2,Ho_blg_nlhe_2021,He_NLHE_BI2Se3_2021,Tiwari_NLHE_2021,Min_NLHE_2023,Kang_NLHE_2019,Shvetsov_NLHE_2019,Qin_NLHE_2021}. Unlike the chiral anomaly, the nonlinear Hall response does not require a magnetic field or magnetization, but rather is an intrinsic higher-order response to an electric field~\cite{Du_NLHE_2021_review}.
\\
\indent
Previous theoretical work combined the Boltzmann transport equation (BTE) with Berry curvature from model Hamiltonians and heuristic treatments of electronic scattering to study the chiral anomaly~\cite{son_chiral_2013,Kim_chiral_2014} and NLHE~\cite{Sodemann_NLHE_2015}. These studies have advanced qualitative understanding of topological transport while focusing on models with simplifying approximations such as constant relaxation times and Berry curvature dipole (BCD) from tight-binding. 
\\
\indent 
First-principles calculations using density functional theory (DFT)~\cite{Martin-book} and Wannier functions~\cite{Marzari_mlwf_2012} have enabled quantitative studies of electronic band topology and Berry curvature~\cite{Wang_berry_2006,feng_2016_berry}. In parallel, first-principles calculations of electron-phonon ($e$-ph) interactions and phonon-limited electronic transport~\cite{Zhou2016, Jhalani2017, Lee_2018, Lee_2020, Jhalani-quad, Li2015, Sohier2018, Desai2021, David_sro_2023} have seen rapid progress, including in topological materials~\cite{Desai-na3bi, Margine-taas}. 
However, studies combining these two advances are almost absent, with the exception of recent calculations by Lihm \textit{et al.}~\cite{Lihm_NLHE_2024} predicting an enhancement of NLHE in GeTe due to $e$-ph interactions. First-principles calculations of the chiral anomaly have not been reported, and a method to study topological magnetotransport and the NLHE in a unified framework is still missing. 
\\
\indent 
In this Letter, we show calculations that combine the BTE with first-principles $e$-ph interactions and Berry curvature. Using this method, we study magnetotransport in a prototypical Weyl semimetal, TaAs, and compute the nonlinear Hall response and $e$-ph interaction-renormalized BCDs in three materials $-$ bilayer WTe$_2$ (BL-WTe$_2$), strained monolayer WSe$_2$ (ML-WSe$_2$), and bulk BaMnSb$_2$. 
Our approach allows us to access both classical (Lorentz) and quantum (chiral) contributions to magnetotransport by importance-sampling of Brillouin zone regions with large Berry curvature. 
Our calculations in TaAs predict transport properties in very good agreement with experiments and shed light on the relative magnitudes of classical and chiral contributions as a function of Fermi level. 
Our nonlinear transport results reveal a strong interplay between the $e$-ph interactions and BCD, and predict temperature trends of the NLHE consistent with experiments. The approach described in this work paves the way for quantitative studies of intrinsic ($e$-ph limited) magnetotransport and nonlinear transport properties in quantum materials, advancing the microscopic understanding of transport regimes governed by band topology. 
%
%
\vspace{0.1cm}
\\
\indent
We describe the chiral anomaly starting from the semiclassical equations of motion (EOM). In a material with applied electric and magnetic fields ($\mathbf{E}$ and $\mathbf{B}$, respectively), the coupled EOM for the position ($\mathbf{r}$) and momentum ($\mathbf{k}$) of an electron wave-packet are modified by the Berry curvature $\Omega_{n\mathbf{k}}$~\cite{Sundaram_eom_1999,DiXiao_2010, marder2010condensed}. Keeping terms up to $\mathcal{O}(\Omega_{n\mathbf{k}}$) in the Berry curvature, and neglecting the orbital magnetization contribution, the EOM can be written as (see Supplemental Material (SM) for derivations~\cite{supplemental}):
\begin{equation}\label{eq:eom}
\begin{split}
   \dot{\mathbf{r}} &= \frac{1}{1 + \frac{e}{\hbar}\mathbf{B}\cdot\mathbf{\Omega}_{n\mathbf{k}}}\left[\mathbf{v}_{n\mathbf{k}} + \frac{e}{\hbar} \mathbf{E} \cross \mathbf{\Omega}_{n\mathbf{k}} + \frac{e}{\hbar}(\mathbf{v}_{n\mathbf{k}} \cdot \mathbf{\Omega}_{n\mathbf{k}})\mathbf{B}\right]
    \\
    \dot{\mathbf{k}} &= \frac{(-e/\hbar)}{1 + \frac{e}{\hbar}\mathbf{B}\cdot\mathbf{\Omega}_{n\mathbf{k}}}\left[\mathbf{E} +  \mathbf{v}_{n\mathbf{k}} \cross \mathbf{B} + \frac{e}{\hbar}(\mathbf{E} \cdot \mathbf{B}) \mathbf{\Omega}_{n\mathbf{k}}\right] 
\end{split}
\end{equation} 
where $e$ is the electron charge, $\hbar$ is Planck's constant, and $\mathbf{v}_{n\mathbf{k}}$ is the band velocity for a Bloch state with crystal momentum $\mathbf{k}$ and band index $n$. Using these EOMs, we derive a steady-state linearized BTE that takes into account the Berry curvature contribution~\cite{supplemental}\footnote{For topologically trivial materials, the Berry curvature $\mathbf{\Omega}_{n\mathbf{k}}$ vanishes, and Eq.~\eqref{eq:Linearized_BTE} becomes identical to the BTE with both electric and magnetic fields derived previously in Ref.~\cite{Desai2021}.}: 
\begin{equation} \label{eq:Linearized_BTE}
\begin{split}
\!\!\!\frac{1}{1 + \frac{e}{\hbar}\mathbf{B}\cdot\mathbf{\Omega}_{n\mathbf{k}}}\!\left[\mathbf{v}_{n\mathbf{k}}+\! \frac{e}{\hbar}(\mathbf{v}_{n\mathbf{k}} \cdot \mathbf{\Omega}_{n\mathbf{k}})\mathbf{B} +\frac{e}{\hbar} (\mathbf{v}_{n\mathbf{k}} \cross \mathbf{B})\nabla_{\mathbf{k}}\mathbf{F}_{n\mathbf{k}}\right] \\=
\!\frac{1}{\mathcal{N}_{\mathbf{q}}} \sum_{m,\nu \mathbf{q}} (1 + \frac{e}{\hbar}\mathbf{B}\cdot\mathbf{\Omega}_{m\mathbf{k+q}}) W_{n\mathbf{k},m\mathbf{k+q}}^{\nu \mathbf{q}} (\mathbf{F}_{n\mathbf{k}} - \mathbf{F}_{m\mathbf{k+q}}).
\end{split} 
\end{equation}
where $\mathbf{F}_{n\mathbf{k}}$ measures the change in electronic occupations relative to equilibrium, $f_{n\mathbf{k}} - f_{n\mathbf{k}}^0 \!=\! (\partial f^{0}_{n\mathbf{k}}/ \partial \varepsilon_{n\mathbf{k}})\, e\mathbf{E}\cdot \mathbf{F}_{n\mathbf{k}}$,
and $f_{n\mathbf{k}}^0$ is the Fermi-Dirac distribution~\cite{perturbo_paper}. 
The right-hand side of Eq.~(\ref{eq:Linearized_BTE}) is the $e$-ph collision term, where  $W_{n\mathbf{k},m\mathbf{k+q}}^{\nu \mathbf{q}}$ is the scattering rate from state $\ket{n\mathbf{k}}$ to $\ket{m\mathbf{k+q}}$ due to absorption or emission of a phonon (with wave-vector $\mathbf{q}$ and mode index $\nu$) as defined in~\cite{perturbo_paper}, and $\mathcal{N}_{\mathbf{q}}$ is the number of $\mathbf{q}$-points in the summation. 
The factor $(1 + \frac{e}{\hbar}\mathbf{B}\cdot\mathbf{\Omega}_{m\mathbf{k+q}})$ in Eq.~\eqref{eq:Linearized_BTE} can be viewed as a Berry-phase correction to the electronic density of states~\cite{dixiao_2005}. 
\\
\indent 
After solving Eq.~\eqref{eq:Linearized_BTE} for $\mathbf{F}_{n\mathbf{k}}$, we compute the conductivity tensor~\cite{perturbo_paper}
\begin{equation} \label{eq:Conductivity_tensor}
    \sigma_{ab} = \frac{e^2S}{\mathcal{N}_{\mathbf{k}} V k_{\rm{B}}T}\sum_{n\mathbf{k}} (1 + \frac{e}{\hbar}\mathbf{B}\cdot\mathbf{\Omega}_{n\mathbf{k}}) f_{n\mathbf{k}}^{0}(1-f_{n \mathbf{k}}^{0})\, \dot{\mathbf{r}}_a\mathbf{F}_{n\mathbf{k}}(\mathbf{B})_b,
\end{equation}
where $a$ and $b$ are Cartesian directions while $S$, $V$ and $\mathcal{N}_\mathbf{k}$ are the spin-degeneracy, unit-cell volume, and number of $\mathbf{k}$-points, respectively. Note that the term in $\dot{\mathbf{r}}$ proportional to $\mathbf{E}$ is not included in the conductivity calculation.  
The implicit dependence of $\sigma$ on the Berry curvature gives rise to chiral transport in Weyl semimetals. In particular, the anomalous velocity term in Eq.~(\ref{eq:eom}) causes the conductivity to increase with magnetic field (negative LMR), in contrast with the usual decrease caused by the Lorentz force in simple metals~\cite{Armitage_review_2018}.
\\
\indent
The NLHE can be described within the same theoretical framework. We expand the expression for the current density $\mathbf{J}$ to second order in the applied electric field~\cite{Du_NLHE_2021_review}:
\begin{equation} \label{eq:ohmslaw}
    J_{a} = \sigma_{a b}E_{b} + \chi_{abc}E_{b}E_{c},
\end{equation}
where we sum over repeated indices; $\sigma_{ab}$ is defined above and its transverse (Hall) components are zero in systems with time-reversal symmetry~\cite{suarez2025nonlinear}. Therefore, the leading-order Hall response is given by the second-order term $\chi_{abc}E_{b}E_{c}$. To obtain an expression for the third-rank conductivity tensor $\chi$, we start from the current
\begin{equation}
\label{currentdefinition}
    \boldsymbol{\mathrm{J}} = \frac{-Se}{\mathcal{N}_{\mathbf{k}} V} \sum_{n\mathbf{k}}f_{n\mathbf{k}}\dot{\mathbf{r}}.
\end{equation}
Setting $\mathbf{B}$=0 in Eq.~(\ref{eq:Conductivity_tensor}) and using the above expression for $\mathbf{J}$, we obtain (see SM~\cite{supplemental}):
\begin{equation} \label{eq:secondConductivity_tensor}
    \chi_{abc} = \epsilon_{abd}T_{dc} + \epsilon_{acd}T_{db}
\end{equation}
where $\epsilon_{abc}$ is the Levi-Civita tensor and we define
\begin{equation} \label{eq:secondConductivity_tensor}
    T_{ab} =  \frac{e^3S}{2\mathcal{N}_{\mathbf{k}} V \hbar k_{\rm{B}}T}\sum_{n\mathbf{k}} f_{n\mathbf{k}}^{0}(1-f_{n \mathbf{k}}^{0})\, (\mathbf{\Omega}_{n\mathbf{k}})_a(\mathbf{F}_{n\mathbf{k}})_b.
\end{equation}
%
\indent
This formula reduces the computation of the rank-three tensor $\chi_{abc}$ to the computation of the rank-two tensor $T_{ab}$. Previous work on NLHE used a constant relaxation time approximation~\cite{Sodemann_NLHE_2015}, in which case this tensor becomes $T_{ab} \!=\! \frac{e^3S\tau}{2\hbar^2}D_{ab}$, where $D_{ab}$ is the BCD defined in band theory without interactions~\cite{Sodemann_NLHE_2015}:
\begin{equation}\label{eq:original_bcd}
    D_{ab} = \int \frac{d^{d}\mathbf{k}}{(2\pi)^d} \Omega_{n\mathbf{k},a}\frac{\partial f_{n\mathbf{k}}^{(0)}}{\partial k_{b}}.
\end{equation}
Here, we modify this expression to take into account $e$-ph scattering, and define an $e$-ph renormalized BCD:
\begin{equation}
\begin{split}
\label{eq:bcd_eph}
   D_{ab}^{\mathrm{e-ph}} &= \frac{2\hbar^2}{e^3S\, \langle\tau\rangle} T_{ab}\\ 
    &= \frac{\hbar}{\mathcal{N}_{\mathbf{k}}Vk_{B}T\,\langle\tau\rangle}  \sum_{n\mathbf{k}} f^{0}_{n\mathbf{k}} (1 - f^{0}_{n\mathbf{k}})(\mathbf{\Omega}_{n\mathbf{k}})_{a}(\mathbf{F}_{n\mathbf{k}})_{b}
\end{split}
\end{equation}
where $\langle\tau\rangle$ is the average relaxation time obtained from the state-dependent relaxation times using \mbox{$\langle\tau\rangle^{-1} = \sum_{n\mathbf{k}} (\partial f_{n\mathbf{k}} /\partial \varepsilon_{n\mathbf{k}})\,\tau^{-1}_{n\mathbf{k}}/\sum_{n\mathbf{k}}(\partial f_{n\mathbf{k}}\, /\,\partial \varepsilon_{n\mathbf{k}}) $.} As we show in the following, the $e$-ph renormalization, which is encoded in the BTE solution $\mathbf{F}_{n\mathbf{k}}$, has a significant effect on the BCD and is essential to make accurate predictions of nonlinear Hall response.

\begin{figure*}[t]
\centering 
\includegraphics[width=1.0\textwidth]{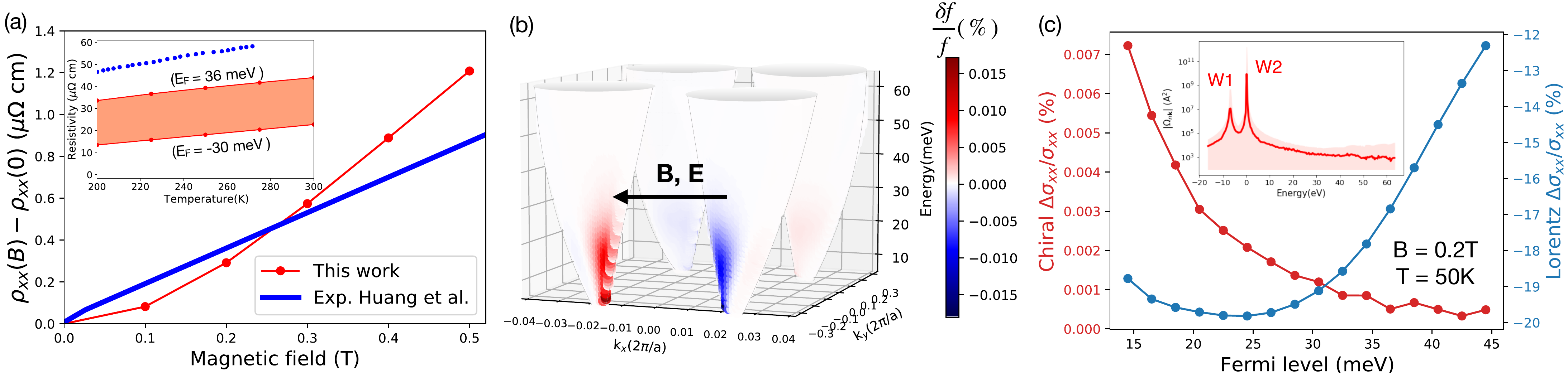}
\vspace{-15pt}
\caption{Magnetotransport in TaAs. (a) Longitudinal magnetoresistance, $\rho_{xx}(\mathbf{B})\!-\!\rho_{xx}(0)$, versus magnetic field, computed at 100~K and compared with experiments from Ref.~\cite{Huang_taas_2015}. The inset shows the resistivity versus temperature for Fermi levels between roughly $\pm$35~meV of the Weyl nodes.  
(b) Weyl cones in TaAs, color-coded according to the change in electronic occupations from the chiral term under parallel electric and magnetic fields, computed at 50~K for $B=0.2$~T. 
(c) Chiral and Lorentz contributions to magnetoconductance versus Fermi level relative to the Weyl node, using the same settings as in (b). The inset shows the Berry curvature as a function of energy for TaAs, with the Weyl cones labeled as W1 and W2.
}\label{fig:taas_data}
\end{figure*}
\indent
To study these transport phenomena, we implement these equations in the open source code \textsc{Perturbo}~\cite{perturbo_paper}. For each material, the electronic ground state, band structure, Berry curvature, and lattice dynamics are calculated using \textsc{Quantum ESPRESSO}~\cite{Giannozzi_QE_2009,Giannozzi_QE_2017}. We then use \textsc{Perturbo} to compute and interpolate the $e$-ph interactions on dense momentum grids and obtain converged transport properties~\cite{perturbo_paper}; the interpolation uses Wannier functions obtained from \textsc{Wannier90}~\cite{Mostofi_2014}. To compute the conductivity in Eq.~(\ref{eq:Conductivity_tensor}), we develop an adaptive $\mathbf{k}$-point sampling technique that can optimally sample states with a large Berry curvature. 
Additional details are provided in the SM~\cite{supplemental}. 
\\

\indent
We first discuss magnetotransport and chiral anomaly in TaAs. 
Figure~\ref{fig:taas_data}(a) shows the classical (Lorentz) contribution to the LMR, $\rho_{xx}(\mathbf{B}) - \rho_{xx}(0)$, as a function of magnetic field for a Fermi level $E_F=35$~meV above the Weyl node at 50~K temperature. Our computed LMR is in very good agreement with experimental data on TaAs from Ref.~\cite{Huang_taas_2015}. The inset shows the phonon-limited resistivity at zero magnetic field as a function of temperature for $E_F$ values between roughly $\pm$35~meV of the Weyl node. For temperatures above $\sim$200~K, where transport is usually limited by $e$-ph interactions, our predicted resistivity is within 30$\%$ of measured values~\cite{Huang_taas_2015}. These results point to a dominant role for the $e$-ph interactions in controlling the resistivity and magnetotransport in TaAs. 
\\
\indent 
In our calculations, TaAs shows 12 pairs of Weyl nodes, with 8 pairs at zero energy (W2) and 4 pairs at -$10$~meV energy (W1), consistent with previous calculations~\cite{Buckeridge_taas_2016,Huang_taas_2015} and angle-resolved photoemission spectroscopy (ARPES) data~\cite{Yang_taasarpes_2015}. 
Figure~\ref{fig:taas_data}(b) shows two pairs of W2 Weyl cones in TaAs, with colors indicating the computed change in electronic occupations at steady state induced by the magnetic field, ($f(\mathbf{B},\mathbf{E})$-$f(0,\mathbf{E})$)/$f(0,\mathbf{E})$, under parallel electric and magnetic fields. 
This effect is caused by the chiral contribution in the BTE 
$-$ the second term in brackets in Eq.~(\ref{eq:Linearized_BTE}): since the sign of the Berry curvature is opposite for the two Weyl cones, the chiral term in the BTE causes an increase in occupations for one cone and a decrease for the other cone in the pair. This creates a net charge pumping effect between Weyl cones, causing the conductivity to increase with magnetic field in our calculations, consistent with previous predictions~\cite{son_chiral_2013, Kim_chiral_2014}. 
\\
\indent
To quantify the increase in conductivity, we compute the chiral contribution to the longitudinal magnetoconductance, shown in Fig.~\ref{fig:taas_data}(c) (red curve) as a function of Fermi level at 50~K for $B$=0.2~T. 
The chiral contribution is large for Fermi levels near the Weyl cones and decays sharply as $E_F$ increases. Since the chiral anomaly is proportional to the Berry curvature, this decay can be attributed to the rapid decrease of the Berry curvature away from the W2 Weyl cones [inset of Fig.~\ref{fig:taas_data}(c)]. 
Note that the divergence of the Berry curvature near the W2 nodes leads to a breakdown of the semiclassical approximation, which restricts our calculations of chiral and classical contributions to $E_F\gtrsim10$~meV from the Weyl nodes (see SM for discussion~\cite{supplemental}). 
\\
\indent 
The classical (Lorentz) term dominates in this energy window due to the rapid decrease of the chiral contribution. 
This trend is clearly seen by computing the Lorentz contribution to the longitudinal magnetoconductance [blue curve in Fig.~\ref{fig:taas_data}(c)], which is found to be orders of magnitude greater than the chiral contribution in the semiclassical regime (here, $E_F >15$~meV from the Weyl nodes). We conclude that the Lorentz term dominates magnetotransport in TaAs in the semiclassical regime. 
This result can help rationalize the experimental data from Ref.~\cite{Zhang_taas_2016}, where the Fermi level is much closer to the Weyl nodes and thus the chiral contribution is expected to govern magnetotransport. 
Transport calculations in that energy window $-$ within a few meV of the Weyl nodes $-$ will require a quantum treatment of transport beyond the BTE and will be explored in future work. 
\begin{figure*}[!th]
\centering 
\includegraphics[width=1.0\textwidth]{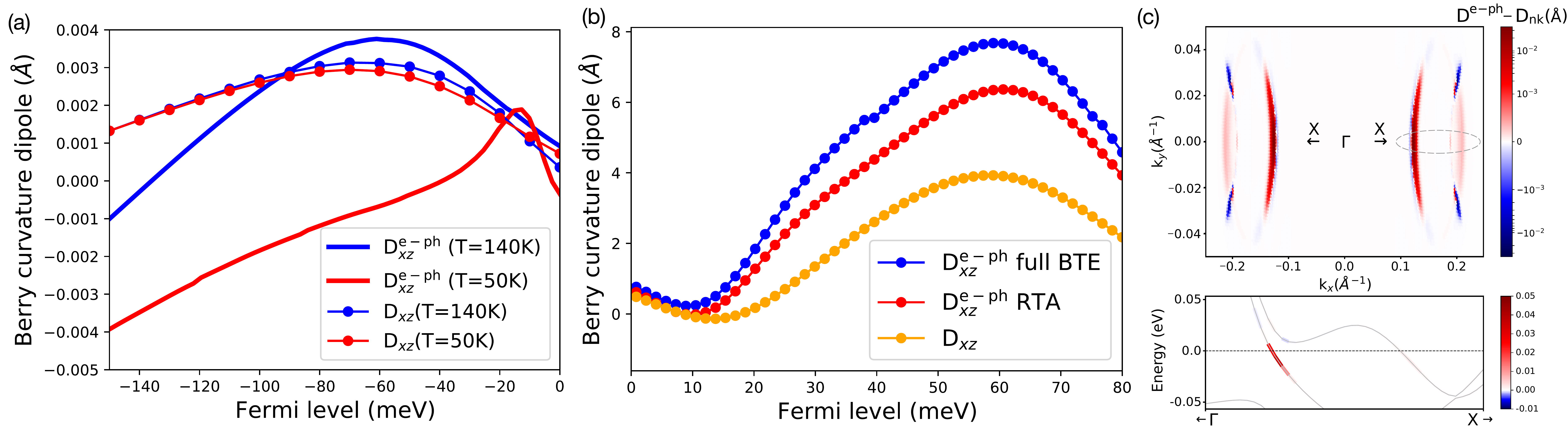}
\vspace{-8pt}
\caption{(a) Berry curvature dipole $D_{xz}$, computed using the Berry curvature alone using Eq.~\eqref{eq:original_bcd}, compared with the $e$-ph renormalized BCD, $D^{\rm e-ph}_{xz}$, in strained ML-WSe$_2$. The results are shown at two temperatures, 50~K (red) and 140~K (blue), as a function of Fermi level in the valence band. (b)  Different approximations for the BCD in BL-WTe$_2$, computed at 100~K without $e$-ph interactions (orange), and with $e$-ph renormalization using the RTA (red) or the full BTE solution (blue). (c) E-ph induced BCD enhancement, shown by plotting the difference of BCD integrands, $D_{n\mathbf{k}}^{\rm e-ph} \!-\! D_{n\mathbf{k}}$, in $\mathbf{k}$-space in BL-WTe$_2$ (upper panel) and the same quantity mapped onto the band structure for the region in the dashed oval (lower panel).}
\label{fig:wse2}
\end{figure*}

Next, we discuss calculations of the NLHE, focusing first on two-dimensional materials, strained ML-WSe$_2$ and BL-WTe$_2$, where previous work has reported the presence of a BCD~\cite{Ma_NLHE_2019,you_dip_2018}. 
In ML-WSe$_2$, it has been shown that breaking the $C_3$ symmetry by applying a uniaxial strain along the $a$-axis induces a finite BCD and Hall response~\cite{you_dip_2018,Qin_NLHE_2021}. Therefore, we conduct our calculations on ML-WSe$_2$ with 3$\%$ strain along the $a$-axis. 
\\
\indent
Addressing the temperature dependence of the BCD and nonlinear Hall response is an important feature of our approach.  
Figure~\ref{fig:wse2}(a) shows the BCD in strained ML-WSe$_2$ computed at two temperatures (50~K and 140~K) for different values of the Fermi level in the valence band. For each calculation, we compare the usual BCD, $D_{xz}$ defined in Eq.~(\ref{eq:original_bcd}) and computed using the Berry curvature alone, with the $e$-ph renormalized BCD, $D^{\rm e-ph}_{xz}$ defined in Eq.~\eqref{eq:bcd_eph}, where $e$-ph scattering is taken into account using the solution of the BTE ($\mathbf{F}_{n\mathbf{k}}$). 
We find that the $e$-ph interactions modify the BCD significantly and change its dependence on temperature and Fermi level [Fig.~\ref{fig:wse2}(a)]. The commonly studied BCD computed directly from the Berry curvature ($D_{xz}$) is nearly constant between 50$-$140~K. In contrast, the $e$-ph interactions introduce a pronounced temperature dependence, and the renormalized BCD increases significantly from 50$-$140~K, consistent with measurements by Qin \textit{et al.}~\cite{Qin_NLHE_2021}.
\\
\indent  
In BL-WTe$_2$, the $e$-ph interactions enhance the BCD but do not qualitatively change the temperature or Fermi-level dependence [Fig.~\ref{fig:wse2}(b)]. 
Figure~\ref{fig:wse2}(b) additionally compares the $e$-ph renormalized BCD ($D^{\rm e-ph}_{xz}$) in BL-WTe$_2$ computed using two different approximations: the \lq\lq full'' solution of the BTE ($\mathbf{F}_{n\mathbf{k}}$ obtained by solving Eq.~\eqref{eq:Linearized_BTE}), which is more accurate and is used throughout this paper, and the simpler relaxation time approximation (RTA), where the form $\mathbf{F}_{n\mathbf{k}} = \tau_{n\mathbf{k}} \mathbf{v}_{n\mathbf{k}}$ is assumed for the deviation of the occupations from equilibrium (note that this version of the RTA still uses first-principles state- and temperature-dependent relaxation times $\tau_{n\mathbf{k}}$, and thus is far more accurate than the \lq\lq textbook'' RTA with an empirical constant relaxation time).   
The renormalized BCD obtained by solving the BTE is 2 times greater than the BCD value without $e$-ph renormalization. The RTA can capture this trend, but it predicts a more modest (50$\%$) enhancement relative to the band theory BCD, with the advantage of bypassing the BTE solution and thus saving significant computational effort.   
\\
\indent
To study the origin of this enhancement, we analyze the contributions to the renormalized BCD by rewriting Eq.~\eqref{eq:bcd_eph} as
\begin{equation}\label{eq:dip_integrand}
\begin{split}
D^{\rm e-ph} \equiv \sum_{n\mathbf{k}} D^{\rm e-ph}_{n\mathbf{k}}\,.
\end{split}
\end{equation}
In Fig.~\ref{fig:wse2}(c), we analyze the effects of $e$-ph interactions by plotting the difference of the $e$-ph-renormalized and band-theory BCD integrands in $\mathbf{k}$-space, $D_{n\mathbf{k}}^{\rm e-ph} - D_{n\mathbf{k}}$, in BL-WTe$_2$. 
We find that the BCD enhancement is concentrated in small regions of the Brillouin zone and arises from electronic states on or near the Weyl cones in correspondence of small electron energy gaps, which enhance the Berry curvature. 
\\
\indent
Using the RTA, $\mathbf{F}_{n\mathbf{k}} = \mathbf{v}_{n\mathbf{k}}\tau_{n\mathbf{k}}$, where $\tau_{n\mathbf{k}}$ is the $e$-ph relaxation time, and thus the renormalized BCD can be written as the band-theory BCD integrand $D_{n\mathbf{k}}$ modified in $\mathbf{k}$-space by the ratio of the state-dependent relaxation time to its average value:
\begin{equation}
D^{\rm e-ph}_{zx,n\mathbf{k}}\label{eq:dip_integrand_orig}
\approx  D_{zx,n\mathbf{k}}\frac{\tau_{n\mathbf{k}}}{\langle\tau\rangle}.
\end{equation}
In BL-WTe$_2$, we have verified that the factor $\tau_{n\mathbf{k}}/\langle\tau\rangle$ is greater than 1 for electronic states with large BCD values, which explains the overall increase in BCD. A similar enhancement is also observed for the BCD obtained from the full BTE solution. The substantial changes in BCD induced by $e$-ph interactions highlight the need for an accurate description of both $e$-ph scattering and band topology for quantitative modeling of nonlinear Hall transport. 

%
%
\begin{figure}[t!]
\centering 
\includegraphics[width=0.8\columnwidth]{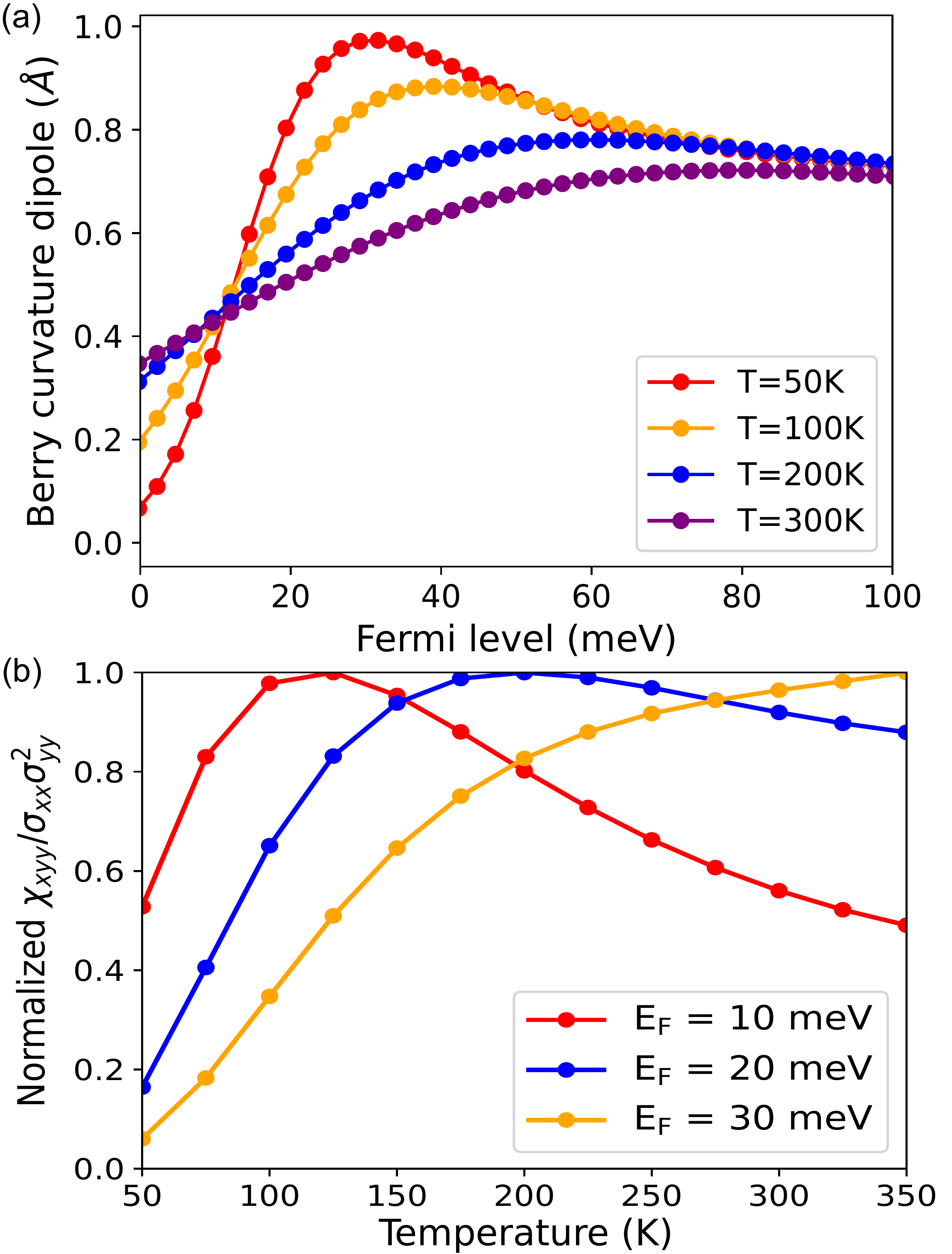}
\vspace{-5 pt}
\caption{Nonlinear Hall effect in BaMnSb$_2$. (a) Berry curvature dipole $D_{yz}$ as a function of Fermi level at four different temperatures. (b) Hall response $\chi_{xyy}/(\sigma_{xx}\sigma_{yy}^2)$ versus temperature, with response peaks normalized to 1 in all cases. Results are shown for three Fermi levels near the conduction band edge, which is taken as the energy zero in both panels.}
\label{fig:bamnsb2}
\end{figure}

Finally, we study the NLHE in BaMnSb$_2$, a bulk Dirac material with a small gap at the Dirac nodes, where a strong nonlinear Hall response near room temperature has been shown in recent experiments~\cite{Min_NLHE_2023}. 
The computed $e$-ph renormalized BCD, shown in Fig.~\ref{fig:bamnsb2}(a), exhibits a pronounced dependence on temperature and Fermi level. 
At low temperature, the BCD shows a sharp characteristic peak for $E_F \!=\! 30$~meV relative to the conduction band edge, consistent with previous calculations~\cite{Min_NLHE_2023}. The nonmonotonic dependence on the Fermi level originates from two competing effects: an increase in phase space for $e$-ph scattering, and a decrease in the Berry curvature, as $E_F$ increases away from the conduction band edge.  
At higher temperatures, the BCD decreases for a wide range of Fermi levels, and the 30-meV peak disappears progressively. We attribute these trends to the increasing strength of $e$-ph interactions at higher temperatures and to the thermal excitation of carriers near the Fermi level, both of which smooth out the 30 meV peak.\\
\indent
A quantity suitable for direct comparison with experiments is the Hall response $V_{xyy}/I_{y}^2$, where $V_{xyy}$ is the nonlinear Hall voltage generated along the $x$-direction when a current $I_{y}$ is applied along the $y$-axis. A simple calculation (see SM~\cite{supplemental}) shows that the nonlinear Hall response measured in experiments is proportional to $\chi_{xyy}/\sigma_{xx}\sigma_{yy}^2$. We compute this quantity for BaMnSb$_2$ in Fig.~\ref{fig:bamnsb2}(b) and show it as a function of temperature for three values of the Fermi level near the BCD peak. The Hall response for $E_{\rm F}$=20~meV shows a peak around 200~K, which is qualitatively consistent with the experimental observation of a nonlinear Hall response peak at high temperature~\cite{Min_NLHE_2023}. 
The position of this peak is sensitive to the Fermi level, and small changes in $E_F$ substantially change the temperature trend. Note also that all responses depend on both second-order conductivities $\chi_{xyy}$ and linear conductivities ($\sigma_{xx}$ and $\sigma_{yy}$). Therefore, calculations of NLHE based only on BCD cannot provide quantitative predictions of the nonlinear Hall response because the scattering time $\tau$ and its temperature dependence are also needed. This further highlights the importance of accurately modeling electron scattering to study the NLHE.  
\\
\indent
In conclusion, this work demonstrates a unified quantitative framework to study magnetotransport and the nonlinear Hall response in quantum materials, advancing their microscopic understanding. 
Our approach combines two pillars of modern first-principles calculations $-$ electronic band structure and wave functions to describe the band topology, and $e$-ph interactions and the Boltzmann equation to model transport properties and temperature dependence. 
We have shown that this method can correctly predict the magnetoresistance in the semiclassical regime in a topological semimetal and resolve its Lorentz and chiral contributions. 
Our analysis of the NLHE in bulk and two-dimensional materials highlights the strong effects of the $e$-ph interactions on the Berry curvature dipole and nonlinear Hall response. %
\\
\indent
The method shown here adds to the first-principles toolkit for studying quantum materials and can be extended in multiple ways. Future work will expand the formalism to include electron-defect scattering at low temperature~\cite{ite-1,Ite_2021} as well as consider Berry curvature multipoles, magnetism, and quantum-metric effects~\cite{Cano2024}.\\ 

%
%
\noindent
D.D. thanks Yao Luo for fruitful discussions. This work was supported by the National Science Foundation under Grant No. OAC-2209262. L.T. is supported by the National Science Foundation Graduate Research Fellowship under Grant No. 2139433. J.J.Z. is supported by the National Natural Science Foundation of China (Grant No. 12104039). J.P. is supported by the Glocal University 30 project (Grant No. 2.0081021.03). This research used resources of the National Energy Research Scientific Computing Center, a DOE Office of Science User Facility supported by the Office of Science of the U.S. Department of Energy under Contract No. DE-AC02-05CH11231 using NERSC award DDR-ERCAP0026831.

\bibliography{sources}
\end{document}